\documentclass[twocolumn,letterpaper,pra,showpacs]{revtex4}
\newcommand{\preprintmargin}{\hspace{0.4\columnwidth}\hspace{-98.4pt}} 
\usepackage{graphicx}
\usepackage{bm}
\usepackage{amsmath}
\begin{document} 
\title{Occupation number and 
fluctuations in the 
finite-temperature Bose-Hubbard model} 
\author{L.\ I.\ Plimak} 
\affiliation{Fachbereich Physik, 
Technische Universit\"at Kaiserslautern, D-67663 Kaiserslautern, Germany.}
\author{M.\ K.\ Olsen} 
\affiliation{Instituto de F\'{\i}sica da Universidade 
Federal Fluminense, Boa Viagem 24210-340, 
Niter\'oi, Rio de Janeiro, Brazil.}
\author{M.\ Fleischhauer} 
\affiliation{Fachbereich Physik, 
Technische Universit\"at Kaiserslautern, D-67663 Kaiserslautern, Germany.}
\date{\today} 
\begin{abstract} 
We study the occupation numbers and number fluctuations of ultra-cold atoms in 
deep optical lattices for finite temperatures within the
Bose-Hubbard model. Simple analytical expressions 
for the mean occupation number and 
number fluctuations are obtained in the weak-hopping regime using 
an interpolation between results from different 
perturbation approaches in the Mott-insulator and superfluid phases.
These analytical results are compared to exact one dimensional numerical calculations 
using a finite temperature variant of the Density-Matrix Renormalisation
Group (DMRG) method and found to have a high degree of accuracy. We also find very good agreement in the
crossover ``thermal'' region. With the present approach the magnitude of
number fluctuations under realistic experimental conditions can be 
estimated and the properties of the finite temperature phase diagram can be studied.
\end{abstract}
\maketitle 

The Bose-Hubbard (BH) model, first studied in the late 1980's 
\cite{Fisher89}, has long been considered a rather academic testing 
ground for analytical as well as numerical approaches in 
quantum statistical mechanics. Its most interesting feature is the
existence of a quantum phase transition \cite{Sachdev} at zero temperature
between a superfluid and a Mott-insulating phase.
With recent advances in the experimental 
techniques of atom trapping and optical lattices, the BH model has 
regained substantial practical importance.
As first noticed in \cite{Jaksch98}, 
the BH Hamiltonian applies to bosonic atoms trapped in a deep  
lattice potential. 
Recent experiments with optical lattices \cite{Greiner02} 
have spectacularly confirmed the existence of the superfluid-insulator transition.
Since the Mott-phase is characterized
by a well defined occupation number of the potential wells,
the phase transition has important potential applications in quantum
information processing \cite{Bloch-QI-exp,Briegel}, 
or Heisenberg-limited matter-wave interferometry 
\cite{Kasevich}, both of  
which require optical lattices with regular filling.
Although techniques to eliminate lattice defects have been designed
\cite{Cirac} an important issue for the implementation of these
applications under experimental conditions are fluctuations in the 
particle number per site at finite $T$. In this work we analyze these fluctuations 
by studying the finite temperature Bose-Hubbard model.

The zero temperature 
BH model has been extensively investigated in the past 
by methods such as the 
Gutzwiller projection ansatz \cite{RokhsarKotliar91}, 
the strong-coupling expansion \cite{FreericksMonien94,FreericksMonien96,ElstnerMonien99A,ElstnerMonien99B} 
and Quantum Monte-Carlo \cite{KrauthTrivedi91,%
BatrouniScalettarZimanyi90,%
Scalettar95,Batrouni02,Svistunov02}, 
as well as with various mean-field approaches 
(see, e.g., \cite{Fisher89,Amico98-00,Stoof01}). A very powerful numerical
technique in the case of one spatial dimension is the
density matrix renormalization group (DMRG) \cite{DMRG}, used to
calculate first and second order 
(i.e., amplitude and number-number) correlations
\cite{KuenerWhiteMonien00}.
Less attention has been paid to the nonzero temperature properties 
of the BH system. We note here that the quantum phase transition exists
in the strict sense only at $T=0$. For finite $T$ the compressibility,
i.e. the change of the particle number per site with the chemical
potential, is always nonzero and thus only an approximate
Mott phase exists. Furthermore the transition between superfluid
and insulating behaviour passes through an intermediate thermal region. 
Recently this thermal crossover has been studied within a slave-boson
formalism \cite{Stoof03}. In this work we study the finite-$T$ properties
using a perturbative analytic approach in the strong-coupling limit, deriving simple analytical expressions for the occupation
number and number fluctuations which cover both quantum and thermal effects. 
To verify these results numerically in one spatial dimension,
we employ a finite-$T$ version of the DMRG
approach \cite{DMRG}. 
We observe good agreement between 
perturbative and DMRG results, including the description
of the thermal crossover between the Mott and insulating phases.

We consider a $d$-dimensional 
cubic lattice of $N$ nonlinear oscillators 
characterised by the following Hamiltonian, 
\begin{align} 
\begin{aligned}
\hat{\cal H} - \mu \hat{\cal N} 
= 
\hat{\cal H}_{\text{hop}} + \hat{\cal H}_0
, 
\end{aligned} 
\label{eq:H} 
\end{align}%
where $\mu $ is the chemical potential, 
$ \hat{\cal N} = \sum_{\bm{k}} \hat n_{\bm{k}}
$
is the operator of the total number of atoms, and 
$\hat n_{\bm{k}} = \hat a_{\bm{k}}^{\dag}\hat a_{\bm{k}}$, 
with $\hat a_{\bm{k}}^{\dag}$, 
$\hat a_{\bm{k}}$ being the usual bosonic 
creation/annihilation pairs. The lattice sites are numbered 
with a $d$-dimensional index ${\bm{k}} = \left\{
k_1, \cdots, k_d
\right\} $, 
while $\sum_{\bm{k}}$ means summation over all lattice sites. 
The first term, 
\begin{equation}
\hat{\cal H}_{\text{hop}} 
= - J \sum_{\langle {\bm{k}}{\bm{l}}\rangle} 
\left(
\hat a_{\bm{k}}^{\dag} 
\hat a_{\bm{l}} 
+ 
\hat a_{\bm{l}}^{\dag} 
\hat a_{\bm{k}} 
\right), 
\label{eq:Hhop} 
\end{equation}
on the RHS of (\ref{eq:H}) 
is the hopping 
Hamiltonian, $J\geq 0$; the sum extends over nearest neighbours and 
we assume cyclic boundary conditions. 
The second term on the RHS of (\ref{eq:H}), 
$
\hat{\cal H}_0 
= 
\sum_{\bm{k}} H_{\text{nl}}\left(
\hat n_{\bm{k}}
\right) 
$, 
is a sum of the nonlinear 
site Hamiltonians
\begin{multline} 
\preprintmargin
H_{\text{nl}}\left(
\hat n_{\bm{k}}
\right) 
= \frac{U}{2} \hat n_{\bm{k}}\left(
\hat n_{\bm{k}} - 1 
\right) - \mu \hat n_{\bm{k}}
\\ 
= \frac{U}{2} \left(
\hat n_{\bm{k}} - \bar n
\right)^2 + \text{const} 
, 
\label{eq:Hnl} 
\preprintmargin
\end{multline}%
where $\bar n = \mu /U + 1/2$ 
is the filling. 
The quantum 
averaging is defined in the standard manner as 
$
\left\langle \cdots\right\rangle = Z^{-1} \text{Tr}
\left[
\text{e}^{-\beta \left(
\hat{\cal H} - \mu \hat{\cal N}
\right) }(\cdots) 
\right] $, where $
Z = \text{Tr}\,\text{e}^{-\beta \left(
\hat{\cal H} - \mu \hat{\cal N}
\right) }
$ 
is the statistical sum, and $\beta =1/T$ is the inverse temperature. 
Note that oscillator units with $\hbar = k_{\text{B}} = 1$ are used throughout this paper. 

While elaborate methods are needed if we wish to exactly pinpoint lobe 
boundaries or calculate long-range correlations, 
local quantities like the average number of 
atoms and the number fluctuations may 
be obtained at lesser cost. 
Let us firstly consider the Mott insulator phase. 
To zero order in hopping, the ground state of the Mott insulator is 
$
\left|0\right\rangle_{J=0} = {\prod_{\bm{k}}} \left|n_0\right\rangle_{\bm{k}}, $ 
where $ n_0 = \textrm{round}(\bar n)
$, cf. Eq.\ (\protect\ref{eq:Hnl}). 
A first order perturbative correction to this state will contain all 
states which are found from $\left|0\right\rangle_{J=0}$ by moving one 
atom to a neighboring site. 
All such ``single-hop'' 
states are eigenstates of ${\cal H}_0$ with 
the same relative energy $U$, resulting in an 
especially simple expression for the ground state to first order in $J$ 
\cite{Burnett03}. 
Up to second order we find 
\begin{align} 
\begin{aligned}
\left|0\right\rangle = \alpha \left(
\openone - \frac{1}{U}{\cal H}_{\text{hop}}
\right) \left|0\right\rangle_{J=0} + {\cal O}\left({\cal H}_{\text{hop}}^2
\right), 
\end{aligned}%
\end{align}%
where $\alpha $ is a normalisation factor.
By direct calculation we then find 
\begin{gather} 
\begin{gathered} 
\left\langle 0\left|\hat n_{\bm{k}}\right|0\right\rangle = n_0, 
\\ 
\delta n^2 = \left\langle 0\left|\hat n_{\bm{k}}^2\right|0\right\rangle - \left\langle 0\left|\hat n_{\bm{k}}\right|0\right\rangle^2 = 
\frac{2 z J^2 n_0(n_0+1)}{U^2} , 
\end{gathered}%
\label{eq:FluctIns} 
\end{gather}%
where $z=2d$ is the number of nearest neighbours in the lattice. 
These formulae are expected to be a good approximation deep in the Mott insulator
phase where a large energy gap exists and the thermal occupation 
of higher levels can be disregarded.

It is worth noting that, if staying within the 
perturbation approach, these expressions cannot be 
extended to the thermodynamic 
limit $N\to\infty$. 
To demonstrate this, consider the formula thus obtained for the 
normalisation constant, $\alpha $:
\begin{gather} 
\begin{gathered} 
\alpha ^2 \left(
1 + \frac{1}{U^2}\left\langle 0\left|{\cal H}^2_{\text{hop}} \right|0\right\rangle
\right) =1 ,
\\
\left\langle 0\left|{\cal H}^2_{\text{hop}} \right|0\right\rangle = z J^2Nn_0(n_0+1). 
\end{gathered} 
\label{eq:ndn2pert} 
\end{gather}%
These results hold only if $z J^2Nn_0(n_0+1) \ll U^2$, therefore 
making the thermodynamic limit impossible. 
(This problem went unnoticed in Ref.\ \cite{Burnett03}.)
On the other hand, local (on-site) quantities should become independent 
of the size of the system when it becomes large enough 
(thermodynamic limit). In fact, our 
DMRG simulations show that Eq.\ (\protect\ref{eq:FluctIns}) equally holds if 
$z J^2Nn_0(n_0+1) \gtrsim U^2$.

In the superfluid phase, one has to account for states 
with the total number of quanta different from $Nn_0$. 
In the limit of zero hopping 
an arbitrary eigenstate is a product state, 
$\left|\vec n\right\rangle = \prod_{\bm{k}} \left|n_{\bm{k}}\right\rangle_{\bm{k}}$. 
With $\bar n$ close to a half-integer, 
it suffices to keep only two states for each site, 
$\left|n_0\right\rangle$ and $\left|n_0+1\right\rangle$, 
effectively turning bosons into 
fermions. We shall call the subspace of 
states with 
$n_{\bm{k}} = n_0, n_0+1$, 
the \textit{f\/}-subspace. 
In this subspace we introduce standard fermionic 
creation and annihilation operators, 
$\hat c_{\bm{k}} \hat c_{\bm{l}} ^{\dag} + 
\hat c_{\bm{l}} ^{\dag}\hat c_{\bm{k}} = \delta _{\bm{k}\bm{l}}$. 
Within the \textit{f\/}-subspace, ($l,m \geq 0$)
\begin{align} 
\begin{aligned}
\big(\hat a _{\bm{k}}^{\dag}\big) ^{l+m}\,\hat a_{\bm{k}} ^l & = 
\big(
c_{\bm{k}}^{\dag}\sqrt{n_0+1}
\big) ^{m} \big(n_0 + \hat c_{\bm{k}}^{\dag} \hat c_{\bm{k}}\big)^l
, 
\\
\big(\hat a _{\bm{k}}^{\dag}\big) ^l \,\hat a_{\bm{k}}^{l+m} & 
= 
\big(n_0 + \hat c_{\bm{k}}^{\dag} \hat c_{\bm{k}}\big)^l \big(
c_{\bm{k}}\sqrt{n_0+1}
\big) ^{m} . 
\end{aligned}  
\end{align}%
Using these relations one can express any bosonic operator 
in fermionic terms. In particular, 
\begin{align} 
\begin{aligned}
H_{\text{nl}} = \frac{U}{2} \left(
\hat c_{\bm{k}}^{\dag} \hat c_{\bm{k}} + n_0 - \bar n
\right)^2 
= \Delta E \, \hat c_{\bm{k}}^{\dag} \hat c_{\bm{k}} +\text{const}
, 
\end{aligned} 
\end{align}%
where $\Delta E = U \left(
{1}/
{2} + n_0 - \bar n
\right)$.
Projection onto the \textit{f\/}-subspace thus makes 
the Hamiltonian linear so it can be directly diagonalised: 
(omitting an additive constant) 
\begin{gather} 
\hat{\cal H} - \mu \hat{\cal N} = 
\sum_{\bm{l}} \left(
E_{\varphi _{l_1}} +\cdots + E_{\varphi _{l_d}} 
\right) \hat f_{\bm{l}}^{\dag} \hat f_{\bm{l}} 
, 
\end{gather}%
where $\varphi _l = 
{2 \pi (l-1)}/
{N} $, $
E_{\varphi } = \Delta E - 2 J(n_0+1)\cos\varphi 
$, and $\hat f_{\bm{l}}$ are related to $\hat c_{\bm{k}}$
by a $d$-dimensional discrete Fourier transform. 
The average on-site particle number and number fluctuations are then 
expressed by the average number of ``fermions'' per site, 
$\Delta n = \left\langle \hat c_{\bm{k}}^{\dag} \hat c_{\bm{k}}\right\rangle$, as 
\begin{align} 
\begin{aligned}
\left\langle \hat n_{\bm{k}}\right\rangle & = 
n_0 + \Delta n , 
\\
\left\langle \hat n_{\bm{k}}^2\right\rangle & = \left\langle \left(
n_0 + \hat c_{\bm{k}}^{\dag} \hat c_{\bm{k}}
\right)^2
 \right\rangle = n_0^2 + (2n_0+1)\Delta n , 
 \\
\delta n^2 & = \left\langle \hat n_{\bm{k}}^2\right\rangle - \left\langle \hat n_{\bm{k}}\right\rangle^2 = \Delta n (1-\Delta n ) 
. 
\end{aligned}
\label{eq:FluctSF} 
\end{align}
For $\Delta n $ we readily find: ($\beta =1/T$) 
\begin{align} 
\begin{aligned}
\Delta n 
= 
\left(
\prod_{s=1}^d
\int_0^{2\pi} 
\frac{d\varphi _s }{2\pi} 
\right) \frac{1}
{1 + \exp \beta 
\sum_{s=1}^dE_{\varphi_s} } \,
. 
\end{aligned} 
\label{eq:Nu} 
\end{align}
This formula holds under the thermodynamic 
limit $N\to \infty$. 

The relations (\ref{eq:FluctIns}) and (\ref{eq:FluctSF}), (\ref{eq:Nu}) are found 
under mutually exclusive conditions. 
Equation (\ref{eq:FluctIns}) follows 
if we assume 
that only the ground state matters, and retain the first non-vanishing 
correction to its wave function; 
with only one level being important 
temperature is of no concern. Conversely, 
equations (\ref{eq:FluctSF}), (\ref{eq:Nu}) 
ignore corrections to the eigenstate wavefunctions 
yet account for energy 
shifts; with many levels being accounted for, thermal properties are retained. 
The question we now address is whether a relation can be derived which covers both insulator and superfluid regions,  
plus, 
even more importantly, the crossover region (which is 
naturally termed the thermal 
region)? 
It would be natural to include perturbative corrections 
to all ``fermionic'' states, but this 
results in some intractable algebra. 
It turns out, however, that for $J\ll U$, simple interpolating 
formulae betwen the two above results may be found
which 
apply not only to the insulator and superfluid regions, 
but also to the thermal crossover region between them. 
Namely, 
\begin{align} 
\begin{aligned}
\left\langle \hat n_{\bm{k}}\right\rangle & = 
n_0 + \Delta n , 
\\
\delta n^2 & = \Delta n (1-\Delta n ) + \frac{2 z J^2 (n_0+\Delta n )(n_0+\Delta n +1)}{U^2} 
, 
\end{aligned}%
\label{eq:Interp} 
\end{align}%
where $\Delta n $ is given by Eq.\ (\protect\ref{eq:Nu}). 
The idea behind (\ref{eq:Interp}) is relatively simple: for small $J$, 
the quantum contribution (\ref{eq:FluctIns}) is negligible in the thermal 
and (even more so) the superfluid regions, while 
in the insulator region, $\Delta n \approx 0$ (or 1), 
so that (\ref{eq:Interp}) 
coincide with (\ref{eq:FluctIns}). 

\begin{figure*} 
\begin{center} 
\includegraphics[width=0.99\textwidth]{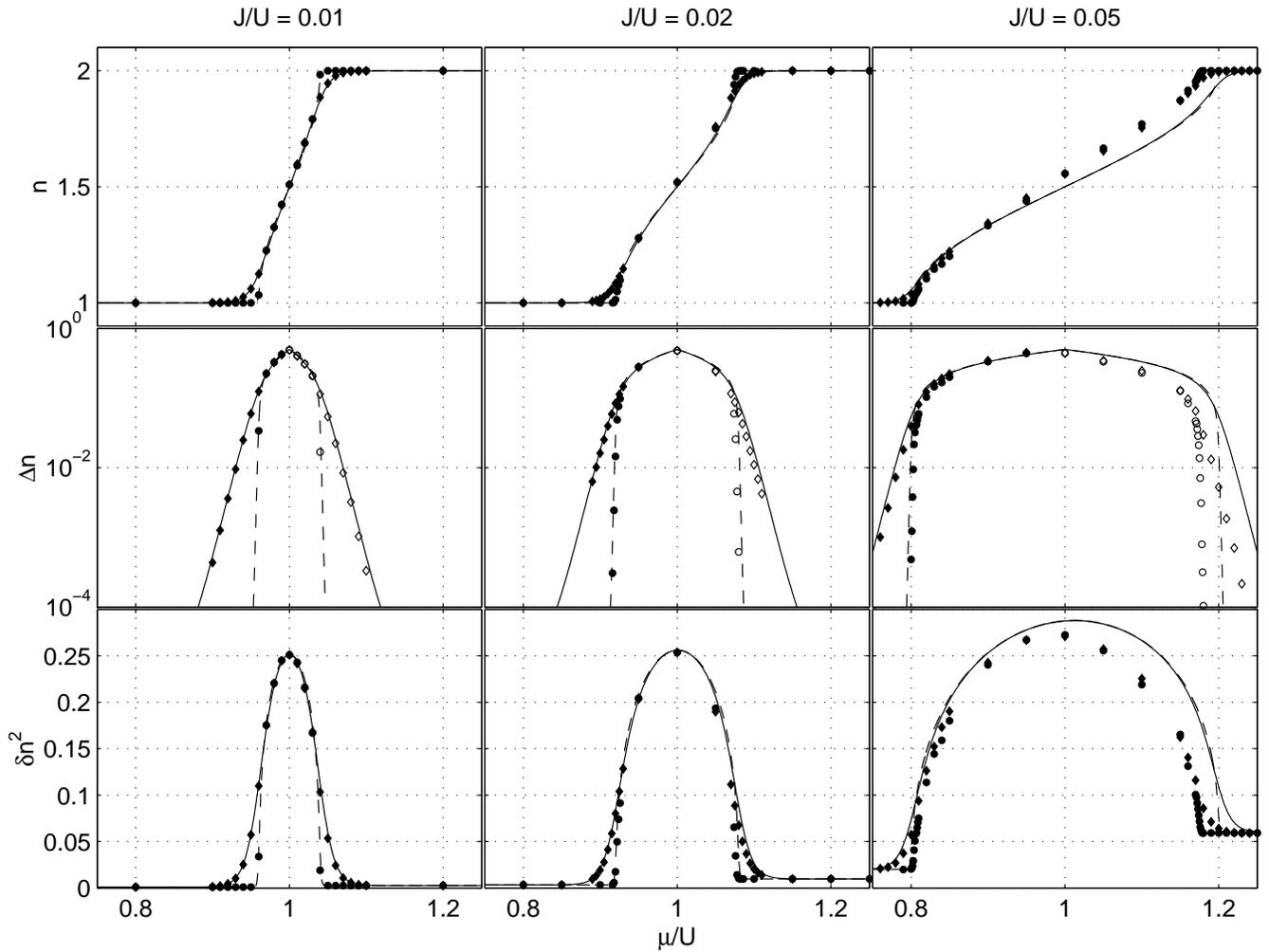}
\end{center} 
\caption{%
Average number and number fluctuations {\em vs\/} $\mu $, 
for two temperatures ($T=0.01U,\ 0.001U$) and three values of the 
hopping parameter ($J = 0.01U,\ 0.02U, \ 0.05U$). 
Top row: on-site population, 
$n = \left\langle \hat n_k\right\rangle$; middle row: difference, $\Delta n$, between 
$\left\langle \hat n\right\rangle$ and the nearest integer; bottom row: on-site 
number fluctuations, $\delta n^2 = \left\langle \hat n_k^2\right\rangle-
\left\langle \hat n_k\right\rangle^2$. 
The lines show perturbative results ($T=0.001U$: solid line, $T=0.01U$: 
dashed line); 
the markers show results of DMRG calculations 
($T=0.001U$: diamonds, $T=0.01U$: circles). 
Open markers are used for $\Delta n<0$.} 
\label{fig:all9}
\end{figure*} 

To verify these results in one spatial dimension, we use a 
finite-temperature version of the 
DMRG approach \cite{DMRG}. 
At the core of our approach is the following 
block-doubling algorithm: 
assuming that we know the optimised states for a block of length $L$, 
we can then find eigenstates of a ring of three such 
blocks, (recall that we use periodic boundary conditions) and calculate the thermal 
$\rho $-matrix. 
Tracing over the states of one block 
yields the $\rho $-matrix of a double-sized 
block. 
Diagonalising this $\rho $-matrix and taking 
a certain number of eigenstates corresponding 
to the largest eigenvalues results in the optimised basis in a block of 
length $2L$. 
The loss of probability is assessed by summing the neglected eigenvalues; 
if this loss becomes unacceptable 
the iterations are stopped. 
To initiate the algorithm, we start from a ring of $L_0+L'$ oscillators, 
and then trace out the states of $L'$ oscillators, resulting in an optimised 
basis for a block of $L_0$ sites. 
We found this to yield much better results than starting from an open-ended 
block by simply selecting its lowest eigenstates. 
The full details of the algorithm will be discussed elsewhere. 

A word of caution is necessary here. 
While borrowing much of the technical 
side of the DMRG method, our approach is conceptually different, 
the difference residing in the importance of temperature. 
Similar to other renormalisation-group methods, DMRG relies on 
the fact that the dimension of the physically relevant subspace 
of the full Hilbert space does not grow with the system
size. For finite temperatures this
can only be true until the thermal correlation length is reached. 
Consequently we only grow the block 
up to a size comparable to the thermal correlation length, and must simply 
stop as soon as 
the probability loss becomes unacceptable. 
We use periodic rather than open boundary conditions based on the same argument. 
For the temperatures considered here the
maximum block length reached in this way is long enough such that
finite-size effects are unimportant. For higher temperatures a combination 
of stochastic and DMRG techniques can be used, which we are developing and will be
discussed in a later work.  

The results of our calculations using both DMRG and perturbative 
approaches in 1-D are summarised in the figure, where 
we plot average on-site numbers 
and number fluctuations 
for three ``cross-sections'' 
along the $\mu $-axis, for $J=0.01U,\ 0.02U,\ 0.05U$. 
Each plot shows data for two temperatures, $T = 0.01U$ and $T = 0.001U$. 
The top and bottom rows of plots represent, respectively, 
the on-site population, 
$\left\langle \hat n_k\right\rangle$, and 
number fluctuations, $\delta n^2 = \left\langle \hat n_k^2\right\rangle-
\left\langle \hat n_k\right\rangle^2$. 
To gain more insight into the thermal region, 
the middle row of plots shows the difference, $\Delta n$, 
between the on-site population 
and the nearest integer value. 
At $T=0.001U$, the insulator region is clearly defined by $\Delta n$
abruptly falling to zero. 
At $T=0.01U$, $\Delta n$ is a smooth function vanishing as $\mu $ 
goes deeper into the insulator 
region. 
A similar effect is seen for the on-site number fluctuations. 
At $T=0.001U$, the phase transition points are well defined, whereas 
at $T=0.01U$ the boundaries of the insulator phase are ``eroded'' 
so that $\delta n^2$ is a visually smooth function of $\mu $.

Comparing the DMRG to perturbative results, we see 
that the latter provide a surprisingly good description of the 
quantities in question. 
For $J=0.01U$ (left column of plots), we find a very good agreement between 
the perturbative and DMRG results. This agreement is reasonable for $J=0.02U$, 
(center column)
becoming only fair at $J=0.05U$ (right column). Note that, 
even in the latter case, the error is mostly in positioning the 
thermal region between the insulator and superfluid phases. 
Away from the thermal region, the results of the perturbative approach show 
good agreement with the DMRG results, both for the superfluid and 
for the insulator. 

\begin{acknowledgments}
This research was supported by 
the Deutsche Forschungsgemeinschaft under contract FL210/12 
and 
the New Zealand Foundation for Research, Science and Technology (Grant No. UFRJ0001). 
\end{acknowledgments}

 
\end{document}